\documentstyle[mprocl,epsf,epsfig]{article}

\def\PRD{{\em Phys. Rev.} D}

\def\be{\begin{equation}}
\def\ee{\end{equation}}
\def\bea{\begin{eqnarray}}
\def\eea{\end{eqnarray}}

\begin{document}

\title{Chaotic exit to inflation: pre-inflationary Friedmann-Robertson-Walker
universes}

\author{G. A. Monerat, H. P. de Oliveira}

\address{Universidade do Estado do Rio de Janeiro, Instituto de F\'\i sica\\
R. S\~ao Francisco Xavier, 524, CEP 20550-013, Rio de Janeiro, RJ, Brazil}

\author{I. Dami\~ao Soares }

\address{Centro Brasileiro de Pesquisas F\'\i sicas\\ Rua Dr. 
Xavier Sigaud, 150, CEP 22290, Rio de Janeiro, RJ, Brazil}

\maketitle\abstracts{We show that homogeneous and isotropic cosmological
models with radiation and scalar field have, in general, two possible chaotic
exits to inflation. These models may describe the early stages of inflation.
The central point of our analysis is based on the possible existence of one or
two saddle-centers points in the phase space of the models.}

\section{Introduction}

In the preceding paper we have examined the role of anisotropy in inducing
chaotic dynamics and chaotic exits to inflation for the Bianchi IX comsologial
models. However, chaotic behavior seems to be a general feature in the early
satges of inflation. We ilustrate this in the realm of rather simple
homogeneous and isotropic cosmological scenario, in which matter is
represented by radiation and scalar fields. Non-integrability and chaos are
invariantly characterized by the topology of homoclinic cylinders\cite{werner}
in the phase space. The relevant manifestation of chaos is that even small
fluctuations of the scalar field are able to preclude or induce the universe
to escape to the inflationay phase with two distinct channels of exit to
inflation.  

\section{The Dynamics of the Model and the Chaotic Exit to Inflation}

The dynamics of closed homogeneous and isotropic cosmological models with
scalar field and radiation is governed by the following Hamiltonian

\begin{equation}
H = -\frac{p_{\varphi}^2}{2\,a^3} + \frac{p_a^2}{12\,a} + 3\,a -
a^3\,V(\varphi) - \frac{E_0}{a} = 0,
\end{equation}

\noindent where $a$, $\varphi$ are the scale factor and the scalar field,
respcetively, whereas $p_a$, $p_{\varphi}$ are the conjugated momenta. From
the conservation of energy and the state equation $p = \frac{1}{3}\,\rho$, it
follows that $\rho = E_0\,a^{-4}$, $E_0$ being a constant of integration.
$V(\varphi)$ is the potential of the scalar field.

There is one or more critical points in the finite region of the phase space
denoted generically by $P$ whose coordinates are $a_0 =
\sqrt{\frac{3}{2\,\Lambda_{ef}}}, \varphi = \varphi_0, p_a = p_{\varphi} = 0$,
where $\varphi_0$ is solution of the equation $V^{\prime}(\varphi_0) = 0$ and
$\Lambda_{ef}=V(\varphi_0)$. The energy associated to the critical point is
$E_0 = E_{cr} = \frac{3}{2}\,a_0^{2}$. Assuming $\Lambda_{ef} > 0$, this
critical point represents the static Einstein universe. The stability analysis
reveals the following eigenvalues associated to $P$: $\lambda_{1,2} =
\pm\,\sqrt{\frac{4}{3}\,\Lambda_{ef}}, \, \, \lambda_{3,4} =
\pm\,\sqrt{-V^{\prime\prime}(\varphi_0)}$. Two possibilities arise: if
$V^{\prime\prime}(\varphi_0) > 0$, $\lambda_{3,4}$ are pure imaginary
indicating that $P$ is a saddle-center\cite{hi}, which induces a cylindrical
topology in the phase space. These eigenvalues are associated to the
hyperbolic and rotational motions around a small neighborhood of $P$. If
$V^{\prime\prime}(\varphi_0) < 0$, $\lambda_{3,4}$ are real with opposite signs
meaning that $P$ is a pure saddle. The phase space also has two critical
points at infinity corresponding to the de Sitter solution, one acting as an
attractor (stable de Sitter configuration) and the other as a repeller
(unstable de Sitter configuration). The scale factor $a(t)$ approaches to the
de Sitter attractor as $a \sim e^{\sqrt{\Lambda_{ef}/3}\,\,t}$, whereas
$p_{\varphi}=0,\varphi=\varphi_0$. Finally, another very important feature of
the resulting dynamical system is the admittance of invariant manifolds
${\cal{M}}$ defined generically by $\varphi = \varphi_0, \,p_{\varphi} = 0$,
where the the dynamics is governed by a two-dimensional system. Such a system
is integrable and its integral curves represent homogeneous and isotropic
universes with radiation and cosmological constant $\Lambda_{ef}$ (cf. Fig.
1).


\begin{figure}[htb]
\begin{minipage}[t]{0.49\textwidth}
\epsfysize=4cm
\epsfxsize=5cm
\epsffile{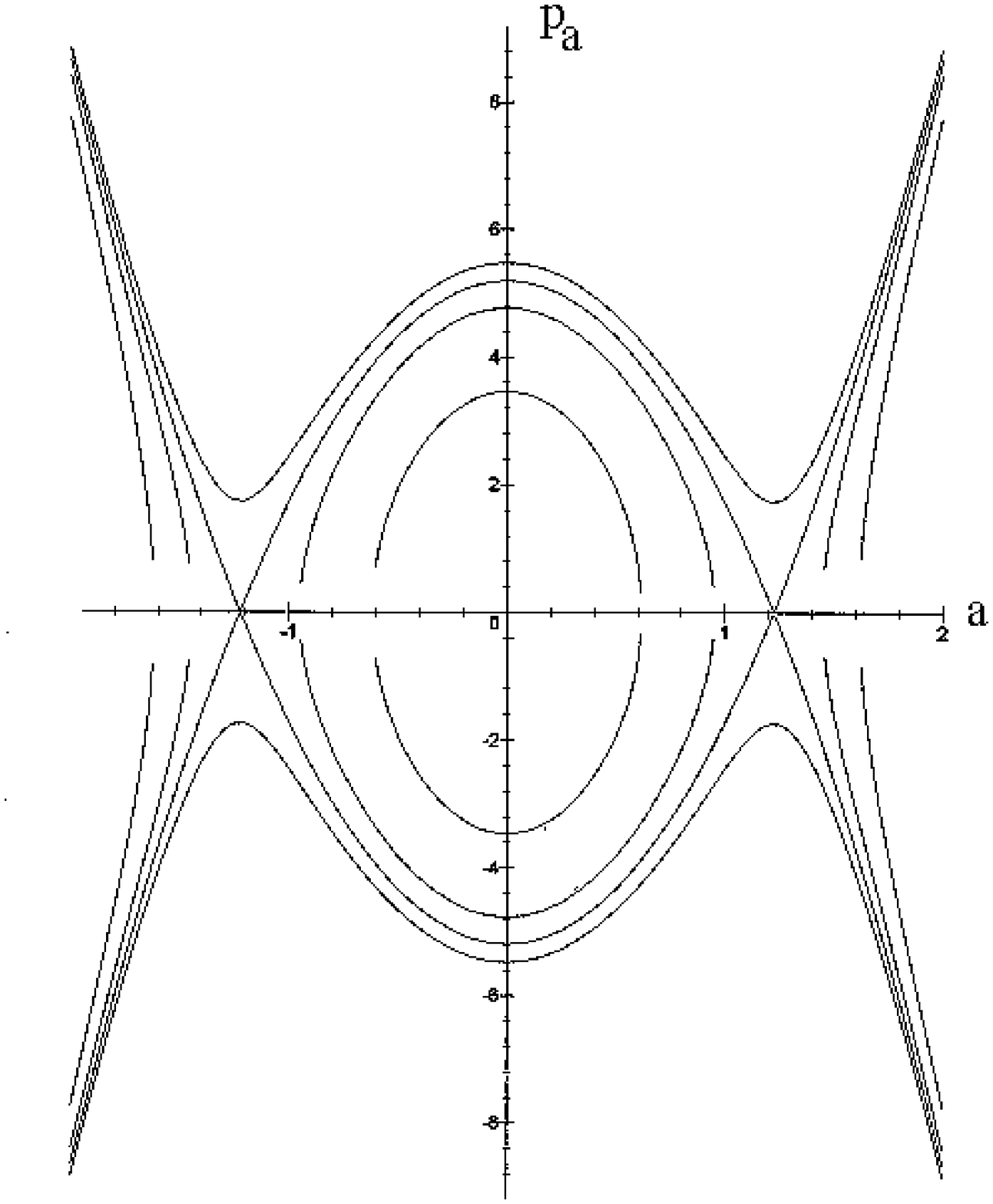}
\caption{Integral curves on the invariant manifold $p_{\varphi}=0, \varphi = {\varphi}_0$ for $\gamma=4/3$ (radiation).}
\end{minipage}
\hspace{0cm}
\begin{minipage}[t]{0.49\textwidth}
\epsfysize=4cm
\epsfxsize=4.5cm
\epsffile{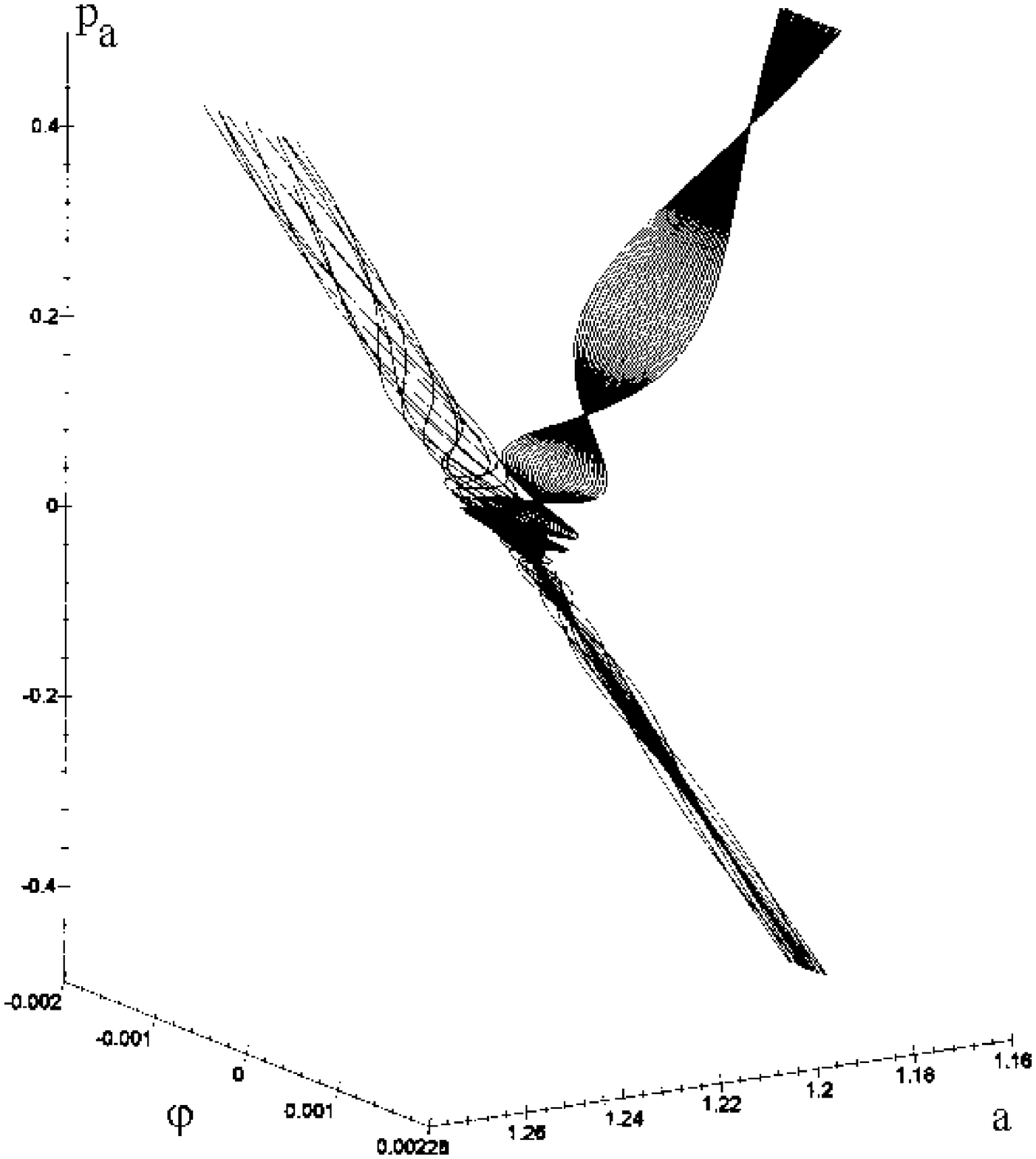}
\caption{Chaotic exit to inflation: 30 orbits taken in a sphere of radius $R=10^{-3}$ about a point $S_0$ on the separatrix.}
\end{minipage}
\end{figure}

The numerical experiments are performed using the package $Poincare$\cite{ed}.
Two distinct situations are considered:$(a)\, \, V(\varphi) = \Lambda +
\frac{1}{2}\,m^2\,\varphi^2$, and $(b)\, \,V(\varphi) = \Lambda +
\frac{\lambda}{4}\,(\varphi^2 - \sigma^2)^2$, where the cosmological constant
$\Lambda$ is included, $m$ is the mass of the scalar field, $\lambda$ and
$\sigma$ are positive constants. For the first choice, the phase space has
only one critical point $P$ identified as a saddle-center. There is also one
invariant manifold ${\cal{M}}$. In the same way, as pointed in Ref. [3], the
phase space under consideration is not compact, and we will actually identify
the chaotic behavior associated to the possible asymptotic outcomes of the
orbits in this phase space, namely, escape to de Sitter state attractor at
infinity (inflationary regime) or collapse after a burst of initial expansion.

Let us consider initially case $(a)$. The behavior of the orbits
near the separatrices that belong to the invariant manifold ${\cal{M}}$ is
performed numerically. We remark, however, that the separatrices define the regions of
collapse and expansion in ${\cal{M}}$, but not in the $full$ 4-dimensional
phase space. Let us consider a point belonging to the separatrix. Around this point,
we construct a 4-dimensional sphere of initial conditions in the phase space
with arbitrary small radius $R$, say $R=10^{-4}$, that represent expanding
models after the initial singularity. The main result is showed in Fig. 2,
where there is always a domain of energy such that part of the orbits escape
to the de Sitter phase and another part collapse, resulting in an
$indeterminate$ outcome. Note that, before the escape or collapse, the orbits
visit a small neighborhood of the saddle-center $P$ from which the partition
of the total energy into rotational and hyperbolic modes about $P$ is
chaotic\cite{hi}. Putting in another words we state that the boundaries of collapse and
escape are chaotically mixed.

The choice of the potential $(b)$ with symmetry breaking offers a richer phase
space dynamics. Three critical points, two of them are saddle-centers whereas
the remaining is a pure saddle, coexist. Also, three invariant manifolds, each
containing one critical point are present. Repeating the numerical experiments
of taking a small sphere of initial conditions very close to the separatrices
(or any other curve) of one of the invariant manifolds for which the critical
point is a saddle-center, we observe the same chaotic behavior with respect to
collapse and escape to the de Sitter phase, as indicated in Fig. 2. However,
if inital conditions are taken near the invariant manifold of the pure saddle,
we note that, depending on the total energy $E_0$, two chaotic escapes to
inflation are possible. In the first, there is always domain of energy $E_0$
such that all orbits evolve towards one of the saddle-centers, perform some
oscillations around it to proceed to collapse or escape to the de Sitter
phase as showed in Fig. 3. In the second, all orbits, after visiting one of
the saddle-centers, return to an infinitesimal neighborhood of the pure
saddle, where part of them collapse and another escape to inflation (see Fig.
4). In this case, the partition of the total energy into the hyperbolic
motions around the pure saddle is completely chaotic. 

\begin{figure}[htb]
\begin{minipage}[t]{0.49\textwidth}
\epsfysize=4cm
\epsfxsize=5cm
\epsffile{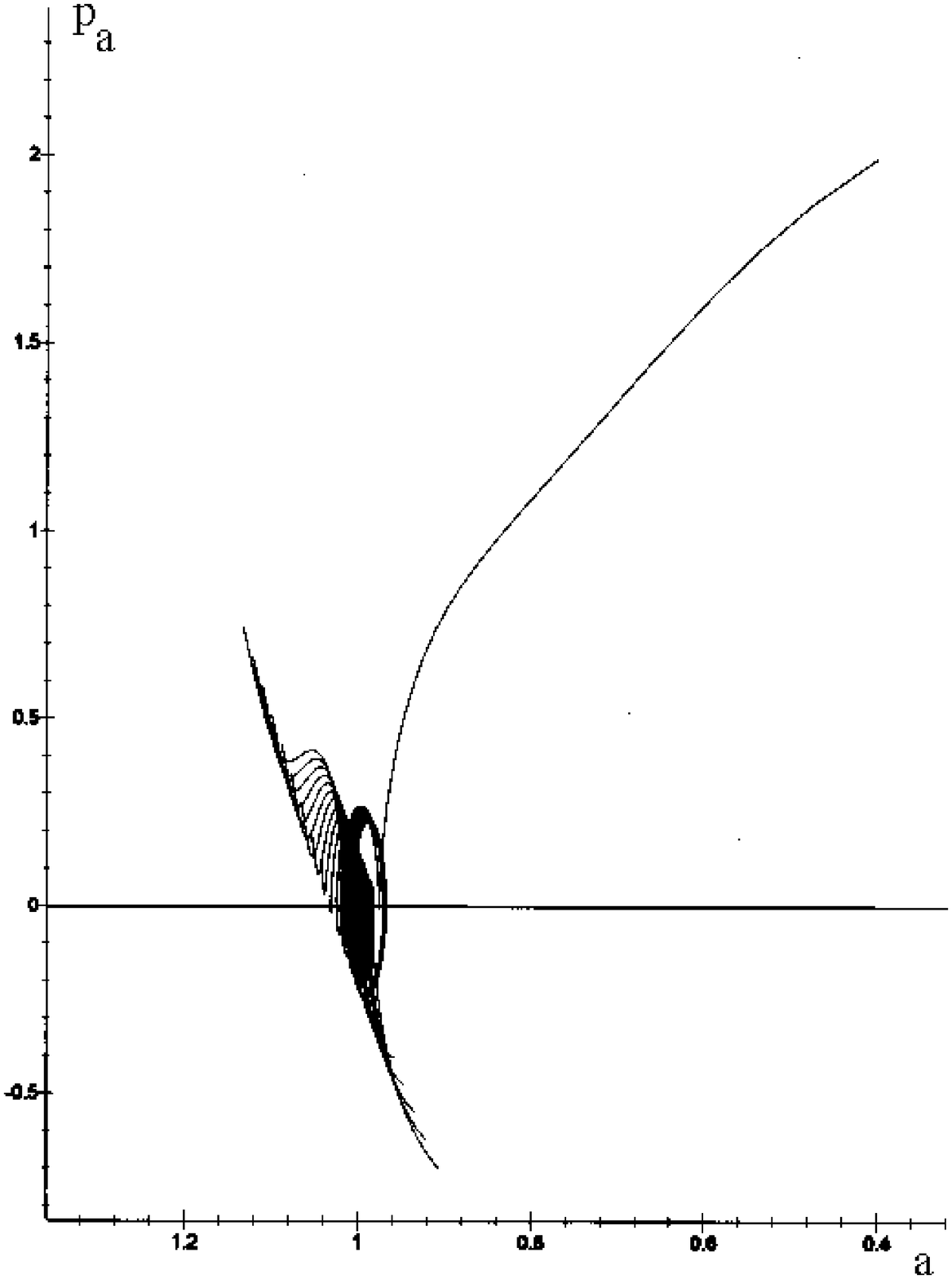}
\caption{Chaotic exit to inflation after oscillations about the saddle-center.}
\end{minipage}
\hspace{0cm}
\begin{minipage}[t]{0.49\textwidth}
\epsfysize=4cm
\epsfxsize=5cm
\epsffile{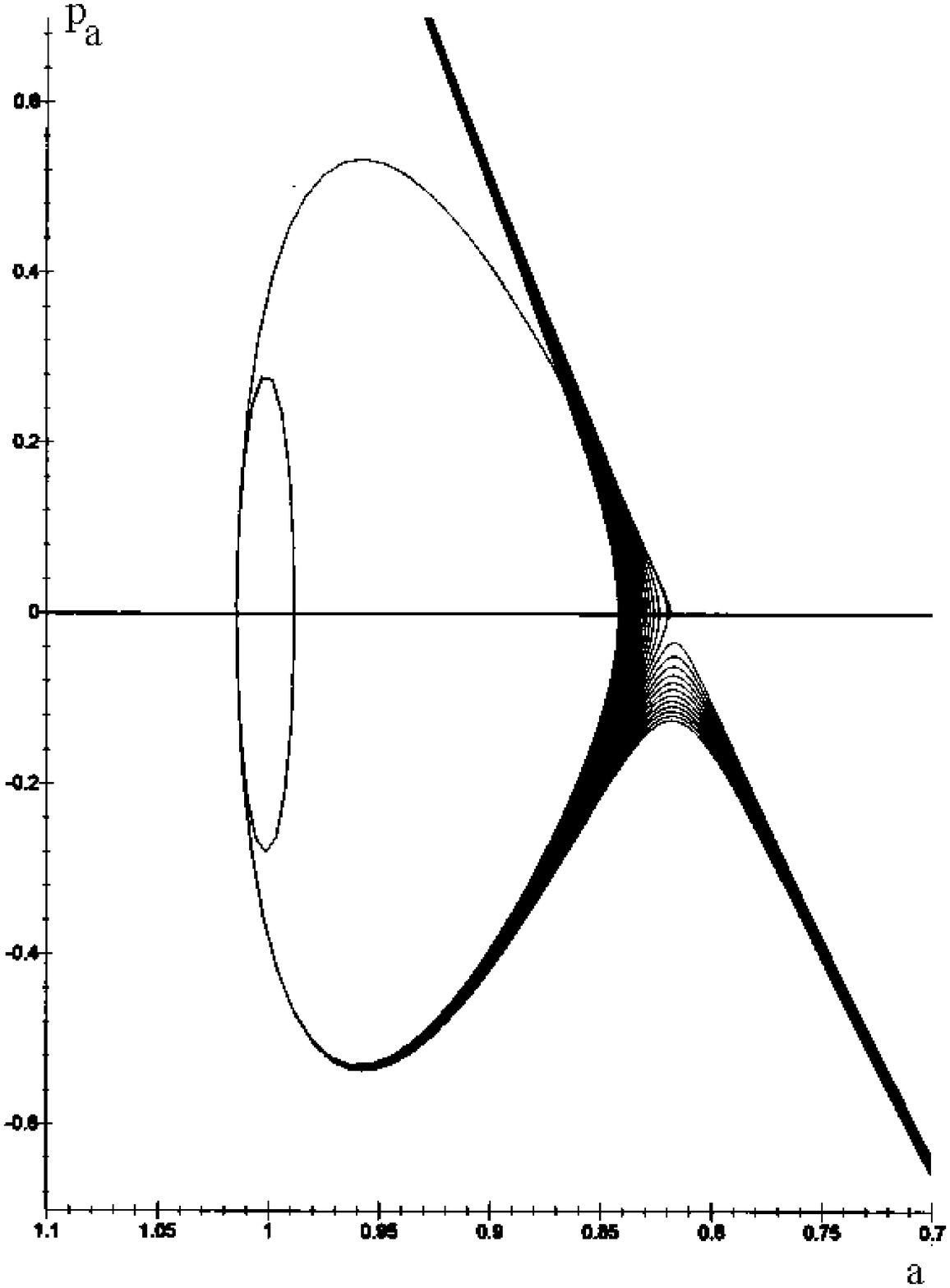}
\caption{Another chaotic exit to inflation: the orbits return to the pure saddle before collapse or escape.}
\end{minipage}
\end{figure}

\vspace{-3mm}

\section*{Acknowledgments}The authors are grateful to CNPq and FAPERJ for
financial support.

\end{document}